# PageRank Pipeline Benchmark:
# Proposal for a Holistic System Benchmark for Big-Data Platforms

Patrick Dreher[1,4], Chansup Byun[2], Chris Hill[3], Vijay Gadepally,[1,2] Bradley Kuszmaul[1], Jeremy Kepner[1,2]

[1]MIT Computer Science & AI Laboratory, Cambridge, MA; [2]MIT Lincoln Laboratory, Lexington, MA;
[3]MIT Department of Earth, Atmospheric and Planetary Sciences, Cambridge, MA;
[4]Department of Computer Science, North Carolina State University, Raleigh, NC
dreher@mit.edu, cbyun@ll.mit.edu, cnh@mit.edu, {gadepally, kepner}@ll.mit.edu, bradley@csail.mit.edu

*Abstract* – The rise of big data systems has created a need for benchmarks to measure and compare the capabilities of these systems. Big data benchmarks present unique scalability challenges. The supercomputing community has wrestled with these challenges for decades and developed methodologies for creating rigorous scalable benchmarks (e.g., HPC Challenge). The proposed PageRank pipeline benchmark employs supercomputing benchmarking methodologies to create a scalable benchmark that is reflective of many real-world big data processing systems. The PageRank pipeline benchmark builds on existing prior scalable benchmarks (Graph500, Sort, and PageRank) to create a holistic benchmark with multiple integrated kernels that can be run together or independently. Each kernel is well defined mathematically and can be implemented in any programming environment. The linear algebraic nature of PageRank makes it well suited to being implemented using the GraphBLAS standard. The computations are simple enough that performance predictions can be made based on simple computing hardware models. The surrounding kernels provide the context for each kernel that allows rigorous definition of both the input and the output for each kernel. Furthermore, since the proposed PageRank pipeline benchmark is scalable in both problem size and hardware, it can be used to measure and quantitatively compare a wide range of present day and future systems. Serial implementations in C++, Python, Python with Pandas, Matlab, Octave, and Julia have been implemented and their single threaded performance has been measured.

*Keywords – benchmarking, big data, supercomputing, PageRank*

## I. INTRODUCTION

Before describing the proposed benchmark we outline, in some detail, the motivation and goals that underlie the benchmark design and scope. Big data processing systems are the backbone of many enterprises. The challenges associated with big data are commonly referred to as the three V's of big data - volume, velocity, and variety [Laney 2001]. Big data volume stresses the storage, memory, and compute capacity of a system and requires access to large amount of computing infrastructure. Big data velocity stresses the rate at which data can be absorbed and meaningful answers produced. Big data variety emphasizes the heterogeneity and dynamic characteristics of information that is processed by big data systems, making it difficult to develop algorithms and tools that can address these diverse data formats.

Many technologies have been developed to address big data volume, velocity, and variety challenges. A typical big data system contains the services shown in Figure 1. A typical big data processing scenario for such a system is as follows. First, data is collected and stored as files. Second, the data is parsed, sorted, and ingested into a database. Third, data from the database is queried and analyzed. Fourth, the results of the analysis are made available to users via web services. The computing resources to run this scenario are brokered by a scheduler on an elastic computing platform.

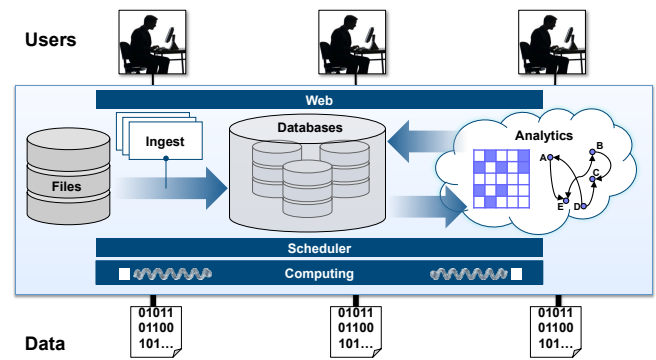

Figure 1. Common architecture for connecting diverse data and users. Typical systems consist of seven major components: files, ingest processes, databases, analytics, web services, and a scheduler that brokers the resources of an elastic computing infrastructure.

The services of a big data processing system can be implemented with a wide range of technologies drawn from both the big data and supercomputing domains.

Common big data software includes the Hadoop distributed file systems (HDFS); Hadoop, Yarn, and Mesos schedulers [Bialecki et al 2005, Vavilapalli et al 2013, Hindman et al 2011]; NoSQL and NewSQL databases including HBase, Accumulo, Cassandra, and SciDB [George 2011, Wall et al 2013, Lakshman & Malik 2010, Balazinska et al 2009]; and analytics environments such as Pig, Hive, Spark, pMatlab, and D4M [Thusoo 2009, Zaharia et al 2010, Kepner 2009, Kepner et al 2012]. These big data technologies are often bundled together by vendors into software stacks that can be deployed onto a system.

Common supercomputing technologies include the parallel file systems such as Lustre and GPFS [Bramm 2004, Schmuck & Haskin 2002]; resource managers/schedulers such as SLURM, LSF, and Maui [Yoo et al 2003, Jackson et al 2001]; and parallel programming environments including MPI, OpenMP, and UPC [Gropp et al 1996, Dagum & Enon 1998, Carlson et al 1999].

Each of the above choices in software can have a significant effect on performance. Big data system builders and technology providers are keenly interested in measuring and understanding the impacts and trade-offs.

Real world big data applications such as text processing, computer network defense, or bioinformatics may perform some or all of these steps in Figure 1. Within a given application, there are many specific operations such as selecting files for further processing or extending a search of a graph to nearest neighbors. Examples of these specific operations are shown in Figure 2. These example operations can be approximately grouped into three categories: initial bulk storage and processing, search and analysis, and administrative tasks.

| Store | Search |
|---|---|
| - Pull data from sources | - Verify permissions |
| - Store data as raw files | - Display query metadata |
| - Select files for further processing | - Collect query logic |
| - Parse files into standard forms | - Collect query arguments/seed |
| - Filter for records of interest | - Form and optimize query |
| - Enrich records with other data | - Execute search |
| - Ingest into database | - Extend search/hop |
| - Correlate data in bulk | - Correlate results, graph analysis |
| - Construct graph relationships | - Summarize results/cluster |
| - Bulk analyze graphs | - Anonymize results |
| Admin | |
| - Create, start, stop, checkpoint, clone, upgrade, restart, … | |

Figure 2. Example operations performed by big data systems divided into three categories: bulk storage and processing, search and analysis, and administrative tasks.

The specific operations listed in Figure 2 are at a sufficient level of detail that it is possible to anticipate which parts of a data processing system (hardware and software) will have the largest impact on the performance of those operations. The different elements of a big data processing system are as diverse as the applications that are performed with these systems and include internal network bandwidth, processor capabilities, memory architectures, database memory usage, implementation languages, and programmer effort. Figure 3 illustrates how specific operations required by a big data application might be impacted by the specific elements of big data system. The application impacts shown in Figure 3 are unique for each application/system combination, so generalizing can be difficult. However, it is often the case that big data systems stress the parts of a system that intensively store and move data around the system.

Qualitative analysis of big data applications, operations, and systems is a useful starting point for assessing big data technologies (hardware and software), but the qualitative analysis must be supplemented with quantitative measurements. Ideally, each real-world big data application could be carefully measured against each big data technology, but this is cost prohibitive. Benchmarks can play a role in informing this discussion by allowing big data technology providers, big data application developers, and big data users to have a common point-of-reference for comparing the capabilities of their systems. Benchmarks do not eliminate the need for each stakeholder to analyze and understand their requirements in detail. Benchmarks do allow this analysis to be spread out among the stakeholders and allow each stakeholder to focus on analyzing what they know best.

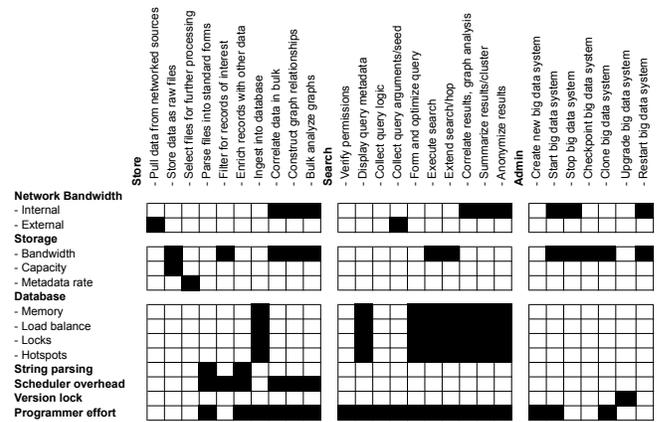

Figure 3. Example analysis with black squares in the table showing the connections between the performance impact of specific elements of a big data system and the operations that the system is performing. Such an analysis is unique to each specific application/system combination.

The major purpose of our proposed benchmark is to create a tool to efficiently inform this type of discussion in the big data space. The design of the benchmark addresses a delicate trade-off between complexity, simplicity, generality, and specificity.

We describe a new big data benchmark rooted in the widely used PageRank algorithm. The benchmark draws heavily on prior benchmarking work that has proved valuable. Section II reviews existing big data and supercomputing benchmarks that have informed the development of the proposed PageRank pipeline benchmark. Section III summarizes PageRank. Section IV describes the PageRank pipeline benchmark. Section V gives a discussion of next steps and future work.

II. SELECTED RELEVANT BENCHMARKS

The rise of big data has resulted in a corresponding rise in big data benchmarks. It is not possible to survey all the relevant big data benchmarks and thus only a few representative big data benchmarks are discussed.

Some important big data benchmarks are the HiBench suite [Huang et al 2010], the Yahoo cloud serving benchmark (YCSB) [Cooper et al 2010], the Big Data Benchmark [Pavlo et al 2009], and Sort [Gray 1988]. The most common element of these benchmarks is their focus on data intensive operations. Most of the computations in the above benchmarks do a relatively small number of operations per data element.

Some important supercomputing benchmarks include Top500 (Linpack) [Dongarra 1988], NAS [Bailey et al 1991], HPC Challenge (Linpack, Stream, FFT, RandomAccess) [Luszczek et al 2006], and Graph500 (Graph Analysis [Bader et al 2007], BFS) [Murphy et al 2010], and HPCG (conjugate gradient) [Dongarra & Heroux 2013]. The most common elements of these benchmarks are their mathematical simplicity and their focus on scalability both in data and in hardware resources. Mathematical simplicity allows the performance of the benchmark to be estimated using simple models of the hardware, which is vital in validating the peak performance of a system. Scalability allows the benchmark to be relevant to a wide range of systems and stand the test of time.

All of these benchmarks (as well as others) can be divided into different categories: goal-oriented, algorithm-oriented, code-oriented, and standards-oriented.

*Goal-oriented benchmarks* specify the required inputs and outputs of the benchmark and usually provide an example algorithm and/or implementation. The user can implement the benchmark with the algorithm/software/hardware combination of their choice. Examples of this approach include NAS, Sort, and Graph500. Goal-oriented benchmarks encourage algorithm/software/hardware co-design and provide a mechanism for highlighting these innovations. In exchange, it is sometimes less clear what part of the system is being measured and it can be difficult for others to correlate benchmark performance with particular application performance.

*Algorithm-oriented benchmarks* specify the required inputs, outputs, and algorithm of the benchmark and provide an example implementation. The user can implement the benchmark with the software/hardware combination of their choice. Examples of this approach include Top500, HPC Challenge (optimized), and HPCG. Algorithm-oriented benchmarks encourage software/hardware co-design and provide a mechanism for highlighting these innovations, and it is usually clear what part of the system is being measured. Algorithm-oriented benchmarks usually allow for proprietary optimizations, and users may not see the same performance without these optimizations.

*Code-oriented benchmarks* provide a specific program that must be run. The user runs the provided code on their choice of system. Examples of this approach include SPEC (spec.org), IOzone (iozone.org), and Intel HiBench. Code-oriented benchmarks provide a mechanism for highlighting hardware innovations and compiler/hardware co-design. Code-oriented benchmarks are easy to produce and easy-to-run, but they usually cannot be used to assess a complete system stack.

*Standards-oriented benchmarks* provide a specific program that must run using specific standard libraries (e.g., MPI, BLAS). The user can implement the benchmark with the libraries/hardware of their choice. Examples of this approach include HPC Challenge (reference) and TPC-C (tpc.org). Standards-oriented benchmarks encourage library/hardware co-design and provide a mechanism for highlighting these innovations. In addition, standards-oriented benchmarks provide a strong incentive for optimizing standard libraries that can be of benefit to a wide class of applications.

The performance of a big data system is strongly influenced by the software environment on the system. Big data benchmarks should be amenable to implementations in diverse environments. Thus, in the big data domain, algorithm-oriented benchmarks would appear to be the most beneficial.

Based on these prior benchmark efforts there are certain properties that are desirable to have in a big data benchmark. These properties include a focus on data intensive operations, mathematical simplicity, and scalability. Existing data intensive benchmarks that satisfy some of these properties include Graph500, Sort (included in HiBench), and PageRank (included in HiBench).

### III. PAGERANK ALGORITHM

PageRank is a link analysis algorithm developed by Google co-founders Sergei Brin and Larry Page [Brin & Page 1998, Page et al 1999]. The algorithm was originally applied rank Web pages for keyword searches. The algorithm measures each Web page's relative importance by assigning a numerical rating from the most important to the least important page within the set of identified Web pages. The PageRank algorithm analyzes the topology of a graph representing links among Web pages and outputs a probability distribution used to represent the likelihood that a person randomly clicking on links will arrive at any particular page.

This algorithm was originally applied specifically to rank webpages within a Google search. However, the mathematics can be applied to any graph or network [Gleich 2015]. The algorithm is applicable to social network analysis [Java 2007, Kwak et al 2009], recommender systems [Song et al 2012], biology [Morrison et al 2005], chemistry [Mooney et al 2012], and neuroscience [Zuo et al 2011]. In chemistry, this algorithm is used in conjunction with molecular dynamics simulations that provides geometric locations for a solute in water. The graph contains edges between the water molecules and can be used to calculate whether the hydrogen bond potential can act as a solvent. In neuroscience, the brain represents a highly complex vertex/edge graph. PageRank has recently been applied to evaluate the importance of brain regions given observed correlations of brain activity. In network analysis PageRank can analyze complex networks and sub-networks that can reveal behavior that could not be discerned by traditional methods.

The simplicity and generality of this algorithm makes it a good candidate for use in a big data benchmark. By judiciously constructing data sets from a graph generator and then adding an ordered set of kernels consisting of file reads, writes, sorts and shuffles, one can construct a data pipeline flow similar to what is required of real world big data systems.

## IV. PageRank Pipeline Benchmark

In many existing HPC micro benchmarks the extract, transform and load operations are often not fully considered when designing big graph and big data implementations. As a result, the cost of these operations is not fully recognized in many benchmark implementations. Nevertheless, they are important components in determining performance and this proposed benchmark addresses these often neglected operations.

The proposed PageRank Pipeline benchmark consists of four mathematically defined kernels that culminate with performing the PageRank algorithm as defined by PageRank on Wikipedia [Wikipedia 2015]. The kernels consist of kernel 0 generating a graph and writing it to files; kernel 1 reading in the files, sorting by the starting vertex and writing out again; kernel 2 reading in the edges, constructing an adjacency matrix, computing the in-degree, eliminating high/low degree nodes, and normalizing each row by total number of edges in the row; kernel 3 computing 20 iterations of PageRank via a sparse matrix vector multiply. The linear algebraic nature of PageRank makes it well suited to being implemented using the GraphBLAS standard. Broadly kernels 0-1 characterize canonical ingest processes (see Figure 1), while kernels 2-3 are akin to canonical analytics stages (see Figure 1). Each kernel in the pipeline *must* be fully completed before the next kernel can begin. Details of the individual kernels in the benchmark are as follows.

### A. Kernel 0: Generate Graph

Kernel 0 generates a list of edges from an approximately power-law graph using the Graph500 graph generator (i.e., kernel 0 of Graph500). Matlab/Octave code for the generator can be obtained from the Graph500 website (Graph500.org). The parameters of the Graph500 generator are the integer scale factor S and the average number of edges per vertex k=16. The maximum vertex label is given by

$$N = 2^S$$

The total number of edges is given by

$$M = k\,N$$

Thus, for a value of S = 30, N = 1,073,741,824, and M = 17,179,869,184. A target scale for the benchmark could be a value of S that results in the memory footprint of the edge data consuming ~25% of the available RAM.

The Graph500 generator is scalable, can be run in parallel without requiring communication between processors, and has been used to generate some of the largest graphs in the world [Burkhardt & Waring 2015, Kepner et al 2014]. Each edge in the graph is defined by a pair of numbers representing the start and end vertices of the edge. For example, let all the starting and ending vertices be stored in the M element vectors **u** and **v**. After the edges are generated they are *written* to files on *non-volatile storage* as pairs of tab separated numeric strings with a newline between each edge:

$$\begin{array}{cc} \mathbf{u}(1) & \mathbf{v}(1) \\ \vdots & \vdots \\ \mathbf{u}(i) & \mathbf{v}(i) \\ \vdots & \vdots \\ \mathbf{u}(M) & \mathbf{v}(M) \end{array}$$

where i = 1, ..., M. The number of files is a free parameter to be set by the implementer or the user. The graph generation process is untimed and its performance is not part of the benchmark.

The Graph500 generator has been a highly successful generator. The subsequent kernels should be able to work with input from any graph generator. Other generators also exist such as block two-level Erdos-Rényi (BTER) [Seshadhri et al 2012] and perfect power law (PPL) [Kepner 2012, Gadepally 2015]. These graph generators may be worth investigating as they may make the validation of subsequent kernels easier.

### B. Kernel 1: Sort

Kernel 1 reads in the files generated in kernel 0, sorts the edges by start vertex and *writes* the sorted edges to files on *non-volatile storage* using the same format as in kernel 0:

$$\begin{array}{cc} \mathbf{u}(1) & \mathbf{v}(1) \\ \vdots & \vdots \\ \mathbf{u}(i) & \mathbf{v}(i) \\ \vdots & \vdots \\ \mathbf{u}(M) & \mathbf{v}(M) \end{array}$$

where $\mathbf{u}(i-1) \leq \mathbf{u}(i) \leq \mathbf{u}(i+1)$.

The number of files is a free parameter to be set by the implementer or the user. The entire sorting process is timed and its performance is reported in terms of edges sorted per second (i.e., M divided by the run time). This kernel has many similarities to the Sort benchmark and its performance should be similar and be dominated by a combination of the storage I/O time and the communication required to sort the data. The type of sorting algorithm may depend upon the scale parameter. For example, in the case where **u** and **v** fit into the RAM of the system, an in-memory algorithm could be used. Likewise, if **u** and **v** are too large to fit in memory, then an out-of-core algorithm would be required.

*C. Kernel 2: Filter*

Kernel 2 reads in the files generated in kernel 1 and performs several filtering steps that are common for preparing a graph for subsequent analysis. The steps are described below along with their Matlab/Octave equivalents.

The first step consists of creating an N x N sparse adjacency matrix of the graph

$$\mathbf{A} = \text{sparse}(\mathbf{u},\mathbf{v},1,N,N)$$

where **A**(u,v) is the count of edges starting at vertex u and ending at vertex v. The matrix construction stores a count at each entry because a (u,v) edge may be generated during kernel 0 more than once. Because of collisions, **A** should have fewer than M non-zero entries, but all the entries in **A** should sum to M. Many rows and columns of **A** may be empty. Many entries along the diagonal of **A** are also expected. Because of the deterministic nature of the PageRank algorithm, none of these factors should significantly impact the run-time of the benchmark.

The second step in kernel 2 is to compute the in-degree of each vertex (i.e., the sum of entries in each column)

$$\mathbf{d}_{in} = \text{sum}(\mathbf{A},1)$$

The third step is to zero-out the columns with the most entries (i.e., eliminating the super-node) and zero-out the columns with only one entry (i.e., eliminating the leaves)

$$\mathbf{A}(:,\mathbf{d}_{in} == \max(\mathbf{d}_{in})) = 0$$
$$\mathbf{A}(:,\mathbf{d}_{in} == 1) = 0$$

The fourth step is to compute the out-degree of each vertex (i.e., number of entries in each row) and divide each *non-zero entry* by its out-degree.

$$\mathbf{d}_{out} = \text{sum}(\mathbf{A},2)$$
$$\mathbf{i} = \mathbf{d}_{out} > 0$$
$$\mathbf{A}(\mathbf{i},:) = \mathbf{A}(\mathbf{i},:) ./ \mathbf{d}_{out}(\mathbf{i})$$

The entire process to perform all of these steps is timed, and its performance is reported in terms of edges prepared per second (i.e., M divided by the run time). In a parallel implementation, a common decomposition would be to have each processor hold a set of rows, since this would correspond to how the files have been sorted in kernel 1. In such a decomposition, the in-degree info will need to be aggregated and the selected vertices for elimination broadcast. This part of this kernel can characterize the relevant network communication capabilities of a big-data system. However, it is possible to construct scenarios in which different steps of kernel 2 could be dominant: reading in the edges (IO limited), constructing the sparse adjacency matrix (memory limited), or computing the in-degree (network limited).

It should be noted that in building the adjacency matrix there may be nodes on the graph with no out edges. Various authors [Boldi, et. al. 2007, Langville and Meyer 2004, Govan et. al. 2008] have proposed adjustments to the adjacency matrix to compensate for the appearance of these dangling nodes. However, these initial Kernel 2 specifications have not adjusted for these for these vertices because it is likely to have limited impact on the run time of the benchmark. Future versions of this Kernel may adjust for these vertices.

*D. Kernel 3: PageRank*

Kernel 3 performs 20 iterations of the PageRank algorithm on the normalized adjacency matrix of the graph provided by kernel 2. In a real application, PageRank would be run until the result passes a convergence test such as the normed sum of the differences between iterations. As PageRank has become more used as a benchmark, this data dependent element of the algorithm is been replaced by running PageRank for a fixed number of iterations [Ewen et al 2012, Gonzalez et al 2014, Kyrola et al 2012, Shun & Blelloch 2013, McSherry et al 2015]. Running PageRank with a set number of iterations yields more consistent timing results that are less dependent on the specifics of the data generator.

The PageRank algorithm is initialized by setting the N-element row vector **r** to normalized random values

$$\mathbf{r} = \text{rand}(1,N)$$
$$\mathbf{r} = \mathbf{r} ./ \text{norm}(\mathbf{r},1)$$

An N-element damping vector **a** is constructed by

$$\mathbf{a} = \text{ones}(1,N) .* (1-c) ./ N$$

where $c = 0.85$ is the damping factor associated with the PageRank algorithm (see Appendix). Using the iterative formulation of PageRank, each iteration of the algorithm performs the following update to the vector **r**

$$\mathbf{r} = ((c .* \mathbf{r}) * \mathbf{A}) + (\mathbf{a} .* \text{sum}(\mathbf{r},2))$$

which can be simplifed to

$$\mathbf{r} = ((c .* \mathbf{r}) * \mathbf{A}) + ((1-c) .* \text{sum}(\mathbf{r},2))$$

The appendix discusses this formula in somewhat more detail.

It should be mentioned that in order to assure a full stochastic construction, the additional term for the dangling nodes in the iterative formulation should be included. Several procedures have been proposed [Eiron McCurley and Tomlin, 2004]. Ipsen and Selee [Ipsen and Selee, 2007] have shown the inclusion of dangling nodes in the PageRank calculation does not materially impact the solution for the PageRank vector. Because this paper is focused on a proposed Kernel 3 for benchmarking rather than specifically finding the PageRank vector **r,** the additional term for the dangling nodes in the iterative formulation has been omitted.

The entire process to perform all of these steps is timed and the performance is reported in terms of edges processed per second (i.e., 20M divided by the run time). In a parallel implementation, a common decomposition would be to have each processor hold a set of rows, since this corresponds to how the files are sorted in kernel 1. In such a decomposition, each processor would compute its own value of **r** that would be summed across all processors and broadcast back to every processor. This is likely to be a time consuming part of this step and is likely to be limited by network communication.

The results of the above calculation can be checked by comparing **r** with the first eigenvector of

$$c.*\mathbf{A}.' + (1 - c)/N$$

For small enough problems where the above dense matrix fits into memory, the first eigenvector can be computed via the Matlab command

$$[\mathbf{r}_1 \sim] = \text{eigs}(c.*\mathbf{A}.' + (1 - c)/N, 1)$$

Normalizing both **r** and $\mathbf{r}_1$ by the sums of their absolute values, these quantities should be equivalent and satisfy

$$\mathbf{r}./\text{norm}(\mathbf{r},1) = \mathbf{r}_1./\text{norm}(\mathbf{r}_1,1)$$

## IV. SERIAL IMPLEMENTATIONS AND RESULTS

Each kernel discussed in the previous section is well defined mathematically and can be implemented in any programming environment. To test this proposed PageRank Pipeline benchmark, serial codes have been developed in several different languages. These include versions written in C++, Python, Python with Pandas, Matlab, Octave, and Julia.

Table I shows the source lines of code needed to implement the serial version of the benchmark in each of the various languages. The C++ implementation is the largest. The other implementations are approximately comparable in size.

TABLE I.  SOURCE LINES OF CODE

| Language | Source Lines of Code |
|---|---|
| C++ | 494 |
| Python | 162 |
| Python w/Pandas | 162 |
| Matlab | 102 |
| Octave | 102 |
| Julia | 162 |

Each implementation was run and timed over a variety of problem sizes corresponding to scale factors that ranged from 16 to 22 (see Table II). Scale 22 results in a problem with maximum of 4M vertices, 67M edges, and an approximate memory footprint of 1.6GB (assuming 16 bytes per edge).

TABLE II.  BENCHMARK RUN SIZES.

| Scale | Max Vertices | Max Edges | ~Memory |
|---|---|---|---|
| 16 | 65K | 1M | 25MB |
| 17 | 131K | 2M | 50MB |
| 18 | 262K | 4M | 100MB |
| 19 | 524K | 8M | 201MB |
| 20 | 1M | 16M | 402MB |
| 21 | 2M | 33M | 805MB |
| 22 | 4M | 67M | 1.6GB |

All of the serial versions of the benchmarks were run on the same hardware architecture and storage environment. The computer platform used was an Intel Xeon E5-2650 (2 GHz) with 64 Gbytes of memory. Each node had 16 cores with hyper threading available. The cluster has both InfiniBand and 10 GigE interconnects. However, because these are all written as serial codes run using a single thread, the network hardware and interconnections were not a major factor impacting the results. The storage system attached to the compute platform used for the read/write/store operations is a Lustre file system.

The measurements for Kernel 0 are shown in Figured 4. This measurement provides some insight into the performance of the code for writing data to non-volatile storage. Although for problems of this, caching in unavoidable.

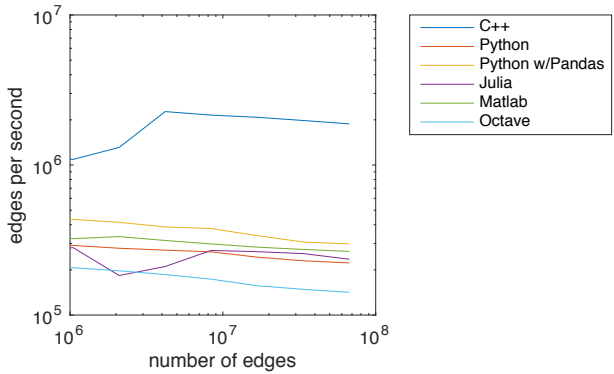

Figure 4. Kernel 0 measurements for each language listed showing edges/sec versus number of edges run on a common hardware platform and file system.

Figure 5 measures Kernel 1. As was discussed in Section 3, if the start and end vertices are sufficiently small, they can fit into memory and an in-memory algorithm can be used. For these measurements, the scale factor of 22 is sufficiently large so as to limit any L3 cache advantage but some impacts on Kernel 1 advantages can still be impacted by the storage cache.

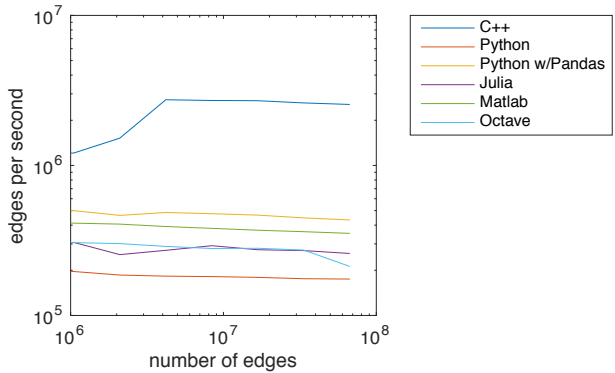

Figure 5. Benchmark results for Kernel 1 showing the performance for reading files generated in Kernel 0, sorting them by the start vertex and re-writing the sorted data back to non-volatile storage.

Kernel 2 measurements are shown in Figure 6. These benchmark tests indicate the impact of I/O, and memory limitations through a combination of reading data, construction of the sparse adjacency matrix and computations to determine the in-degree.

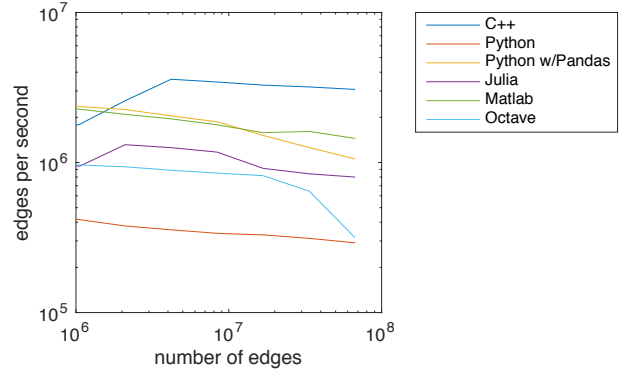

Figure 6. Benchmark results for Kernel 2 showing combined impacts from I/O and memory limitations.

Figure 7 measures the calculation of the actual PageRank algorithm. It should be noted that for this serial implementation, there is a minimal dispersion among the performance measurements in Kernel 3 for each of the languages. This is not be surprising because of the fact that there is no parallel implementation in these tests and therefor there is little network communication. It is expected that measurements of Kernel 3 in a parallel implementation will show a wider dispersion in performance between the languages.

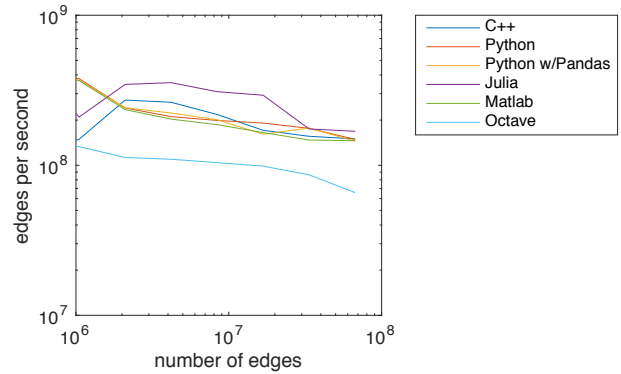

Figure 7. Kernel 3 measurements reflecting the actual PageRank calculations for scales between 16 and 22.

V. SUMMARY AND NEXT STEPS

The proposed PageRank Pipeline benchmark employs supercomputing benchmarking methodologies to create a scalable benchmark that is reflective of many real-world big data processing systems. The PageRank pipeline benchmark leverages existing prior scalable benchmarks (Graph500, Sort, and PageRank) to create a holistic benchmark with multiple integrated kernels that can be run together or independently.

Initial measurements using serial code developed in several difference languages have been presented here using a common hardware platform and a Lustre file system. Future work will include re-running these benchmarks using local storage.

The key next step is to obtain community feedback on the proposed benchmark and make improvements based on that feedback. Possible points of feedback include: Should a more deterministic generator be used in kernel 0 to facilitate validation of all kernels? Should the end vertices in kernel 1 also be sorted? Should a diagonal entry be added to empty rows/columns to allow the PageRank algorithm to converge? Are the values of the adjacency matrix required to be floating point values? What outputs should be recorded to validate correctness?

The computations are also simple enough that performance predictions can be made based on simple hardware models. Additional studies are currently underway that will provide a more detailed analysis of each of the kernels with respect to standard models of parallel computation and communication. The results from these models can be used to predict the performance on current and proposed systems.

Finally, after receiving community input and analyzing the performance models, it would be appropriate to produce an executable specification (i.e., Matlab, Python) and reference implementations in various environments (i.e., C/MPI, Java/Hadoop, Python/Spark). Furthermore, implementations using the GraphBLAS standard would allow enable comparison of the GraphBLAS capabilities with other technologies.

## VI. ACKNOWLEDGMENTS

The authors would like to thank David Bader, Justin Brukardt, Chris Clarke, and Steve Pritchard for their helpful comments.

## APPENDIX

The goal of PageRank is rank vertices in a graph based on how likely a random walker of the graph will be at any particular vertex. The strength of PageRank is that the core random walker concept is very flexible and can be used to incorporate a wide range of contextual information. A variety of specific algorithms have been developed based on this concept [Gliech 2015] with names such as *strongly preferential PageRank*, *weekly preferential PageRank*, and *sink PageRank*. For this benchmark, one of the simpler PageRank algorithms is used. The simplest model says that a random walker will walk to another vertex with equal probability. Such a model can be represented by the following iterative calculation

$$\mathbf{r} = \mathbf{r} * \mathbb{1} ./ N$$

where $\mathbb{1}$ is a NxN matrix of all ones. The above equation will converge to a value of

$$\text{sum}(\mathbf{r},2) ./ N$$

A more sophisticated model increases the probability of randomly walking to a connected vertex and is described by the iterative equation

$$\mathbf{r} = ((c .* \mathbf{r}) * \mathbf{A}) + ((1-c) .* \mathbf{r} * \mathbb{1} ./ N)$$

where **A** is the normalized adjacency matrix of the graph constructed as the output of Kernel 2 and c is the weighting factor that balances between walking to a neighbor vertex versus a random vertex. The above equation simplifies to

$$\mathbf{r} = ((c .* \mathbf{r}) * \mathbf{A}) + ((1-c) .* \text{sum}(\mathbf{r},2) ./ N)$$